\documentclass[twocolumn,superscriptaddress,showpacs,aps,prl]{revtex4}
\usepackage{makeidx}
\usepackage{bm}
\usepackage{graphics}
\usepackage[dvips]{epsfig}
\newcounter{fig}

\begin{document}
\title{Screening in gated bilayer graphene}
\author{L.A. Falkovsky}
\affiliation{L.D. Landau Institute for Theoretical Physics, Moscow
117334, Russia} \affiliation{Institute of the High Pressure
Physics, Troitsk 142190, Russia}

\date{\today}      

\begin{abstract}
The tight-binding model of a graphene bilayer  is used to find the
gap between the conduction and valence bands, as a function of
both the gate voltage and as  the doping by donors or acceptors.
The total Hartree energy is minimized and the equation for the gap
is obtained. This equation for the ratio of the gap to the
chemical potential is  determined  only by the screening constant.
Thus the gap is strictly proportional to the gate voltage or the
carrier concentration in  the absence of donors or acceptors. In
the opposite  case, where the donors or acceptors are present, the
gap demonstrates the asymmetrical behavior on the electron and
hole sides of the gate bias.
\end{abstract}
\maketitle

Bilayer graphene has attracted much interest partly due to the
opening of a tunable gap in its electronic spectrum by  an
external electrostatic field. Such a phenomenon was predicted by
McCann and Fal'ko \cite{McF} and can be observed in optical
studies controlled  by applying a gate bias
\cite{OBS,ZBF,KHM,LHJ,ECNM,NC}. In  Refs. \cite{Mc,MAF}, within
the self-consistent Hartree approximation, the gap was derived  as
a near-linear function of the carrier concentration  injected in
the bilayer by the gate bias.
Recently, this problem was
 numerically considered \cite{GLS} using the density functional theory (DFT)
 and including the external charge doping involved with impurities.
 The DFT calculation
 gives the gap  which is roughly  half of the  gap
 obtained in the Hartree approximation. This disagreement was
 explained in Ref. \cite{GLS} as a result of both the inter- and intralayer
 correlations.

 In this Brief Report, we study this problem within the same
 Hartree approximation as in Refs. \cite{Mc,MAF}, but including the external
 doping. We consider  the case, where the carrier
 concentration in the bilayer is  less than 10$^{13}$ cm$^{-2}$,
  calculating  the  carrier concentration on  both layers.  Then, we minimize the
 total energy of the system and find self-consistently both the chemical potential
 and the gap induced by the gate bias.  Our results completely differ
 from those
 obtained  in Refs. \cite{Mc,MAF} even for the range where the external doping
 is negligible. The dependence of
 the gap on the carrier concentration, i.\,e., on the
 gate voltage,
 exhibits an asymmetry at the electron and hole sides of the gate bias.

\begin{figure}[h]
\resizebox{.25\textwidth}{!}{\includegraphics{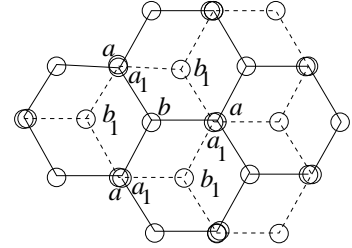}}
\caption{Bilayer lattice} \label{grlat}
\end{figure}

 The  graphene bilayer lattice is shown in Fig. \ref{grlat}.
Atoms  in one layer, i.\,e., $a$ and $b$ in the unit cell, are
connected by  solid lines, and in the other layer, e.\,g., $a_1$
and $b_1$, by the dashed lines. The atom $a$ ($a_1$) differs from
$b$ ($b_1$) because it has a neighbor just below in the adjacent
layer, whereas the atom $b$ ($b_1$) does not.

Let us recall the main results of the Slonchewski--Weiss--McClure
model \cite{SW,McCl}. In the tight-binding model, the Bloch
functions of the bilayer are written in the form
\begin{eqnarray}\nonumber \label{eh}
\psi_a=\frac{1}{\sqrt{N}}\sum_{j}e^{i{\bf ka}_j}\psi_0({\bf
a}_j-{\bf r})\\ \psi_b=\frac{1}{\sqrt{N}}\sum_{j}e^{i{\bf
ka}_j}\psi_0({\bf a}_j+{\bf a}-{\bf r})\\ \nonumber
\psi_{a1}=\frac{1}{\sqrt{N}}\sum_{j}e^{i{\bf ka}_j}\psi_0({\bf a}_j+{\bf c}-{\bf r})\\
\psi_{b1}=\frac{1}{\sqrt{N}}\sum_{j}e^{i{\bf ka}_j}\psi_0({\bf
a}_j+{\bf c}+{\bf a}-{\bf r}),\nonumber
\end{eqnarray}
where the sums are taken over the  lattice vectors ${\bf a}_j$ and
$N$ is the number of unit cells. Vectors  ${\bf a}$ and ${\bf c}$
connect the nearest atoms in the layer and in the neighbor layers,
correspondingly.

 For the nearest neighbors, the effective Hamiltonian in the space
 of the functions (\ref{eh})
contains  the hopping integrals
$\gamma_0,\gamma_1,\gamma_3,\gamma_4,$ and $\Delta$ \cite{PP}. The
largest of them, $\gamma_0$, determines the band dispersion near
the $K$ point in the Brillouin zone with a velocity parameter $v$.
The parameters $\gamma_3$ and $\gamma_4$ giving a correction to
the dispersion are  less than $\gamma_0$ by a factor of 10 (see
Refs. \cite{KHM,LHJ}). The parameters $\gamma_1$ and $\Delta$
result in the displacements of the levels at $K$, but $\Delta$ is
much less than $\gamma_1$. Besides, there is the parameter $U$
induced by the gate  and describing the asymmetry  of two layers
in the external electrostatic field.  This parameter simply
presents the potential energy between two layers, $2U=-edE$, where
$d$ is the interlayer distance and $E$ is the electric field induced
both by the gate voltage and the external charge dopants in the
bilayer. In the simplest case, the effective Hamiltonian can be
written as
\begin{equation}
H(\mathbf{k})=\left(
\begin{array}{cccc}
U \,    & vk_{+} \,& \gamma_1    \, & 0\\
vk_{-} \,& U     \, & 0\,& 0\\
\gamma_1    \,  &0 \,&-U  \,  &vk_{-}\\
0 \,& 0 \,&vk_{+} \,&-U
\end{array}%
\right) ,  \label{ham}
\end{equation}%
where the matrix elements are expanded in the momentum
$k_{\pm}=\mp ik_x-k_y$ near the $K$ points.
\begin{figure}[h]
\resizebox{.4\textwidth}{!}{\includegraphics{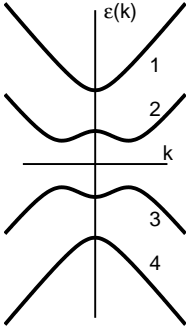}}
\caption{Band structure of bilayer} \label{dis}
\end{figure}

 The Hamiltonian gives four energy bands:
\begin{eqnarray}\label{bands}
\varepsilon_{1,4}(q)=\pm\left(\frac{\gamma_1^2}{2}+U^2+q^2+W\right)^{1/2}\,,\\\nonumber
\varepsilon_{2,3}(q)=\pm\left(\frac{\gamma_1^2}{2}+U^2+q^2-W\right)^{1/2}\,,
\end{eqnarray}
where
$$W=\left(\frac{\gamma_1^4}{4}+(\gamma_1^2+4U^2)q^2\right)^{1/2}$$
and we denote  $q^2=(vk)^2$.

 The band structure is shown in Fig.\,
\ref{dis}. The minimal value of the upper  energy $\varepsilon_1$
is $\sqrt{U^2+\gamma_1^2}$. The
 $\varepsilon_2$ band takes the maximal value $|U|$ at $k=0$ and
the minimal value $\tilde{U}=\gamma_1|U|/\sqrt{\gamma_1^2+4U^2}$
at $q^2=2U^2(\gamma_1^2+2U^2)/(\gamma_1^2+4U^2).$ Because the
value of $U$ is much less than $\gamma_1$, the distinction between
$U$ and $\tilde{U}$ is small and the gap between the bands
$\varepsilon_2$ and $\varepsilon_{3}$ takes the value $2|U|$.

\begin{figure}[h]
\resizebox{.2\textwidth}{!}{\includegraphics{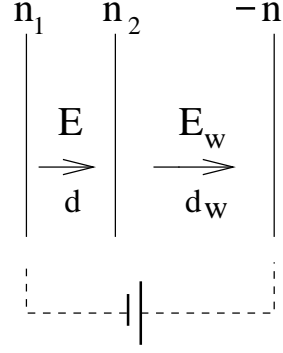}}
\caption{Electrostatic model; $d$ is the interlayer distance,
$d_w$ is the wafer thickness.} \label{bifig1}
\end{figure}

 The eigenfunctions ${\mathbf C}$ of
the Hamiltonian (\ref{ham}) have the form
\begin{equation}
{\mathbf C}=\frac{1}{C}\left(
\begin{array}{c}
(U-\varepsilon_n)[(\varepsilon_n+U)^2-q^2]\\
-q_{-}[(\varepsilon_n+U)^2-q^2]\\
\gamma_1(U^2-\varepsilon_n^2)\\
\gamma_1q_{+}(U-\varepsilon_n)\end{array}\right) ,\label{fun}
\end{equation}
where the ${\mathbf C}$ norm squared is
\begin{eqnarray}\nonumber
C^2=[(\varepsilon_n+U)^2-q^2]^2[(\varepsilon_n-U)^2+q^2]\\\nonumber
+\gamma_1^2(\varepsilon_n-U)^2
[(\varepsilon_n+U)^2+q^2]\,.\end{eqnarray} The probability to find
an electron, for instance, on the first layer is
$|C_1|^2+|C_2|^2$, as seen from Eqs. (\ref{eh}).

We assume, that  carriers  occupy only the bands
$\varepsilon_{2,3}$, so the chemical potential $\mu$ and the gap
$2|U|$ are  less than the distance between the bands
$\varepsilon_1$ and $\varepsilon_2$, i.\,e., $(|\mu|,
2|U|)<\gamma_1.$ The electron dispersion for  the
$\varepsilon_{2,3}$ bands  can be expanded in powers of $q^2$:
\begin{equation}\nonumber
\varepsilon_{2,3}^2(q)=U^2-4\frac{U^2}{\gamma_1^2}q^2+\frac{q^4}{\gamma_1^2}.
\end{equation}
Then, for $q^2\gg 4U^2$, we can use the simple relations:
\begin{eqnarray}\nonumber
\varepsilon _{2,3}(q)=\pm\sqrt{U^2+q^4/\gamma_1^2}\,,\\
|C_1|^2+|C_2|^2=q^4/[q^4+\gamma_1^2(\varepsilon_{2,3}-U)^2]\\
\nonumber =(\varepsilon_{2,3}+U)/2\varepsilon_{2,3}\,.
\end{eqnarray}
Within such the  approximation, many observable effects can be
analytically evaluated for the intermediate carrier concentration,
$4U^2\ll\gamma_1\sqrt{\mu^2-U^2}\ll\gamma_1^2$, where we neglect
the effect of the "mexican hat".

At zero temperature, for the total carrier concentration $n$ and
the carrier concentrations $n_{1,2}$ on the  layers,  we obtain
\begin{eqnarray}\label{n}
n=\frac{\gamma_1}{\pi\hbar^2v^2}\sqrt{\mu^2-U^2}=\frac{n_0U}{\gamma_1}
\sqrt{x^2-1}\,,\\ \label{nn}
n_{1,2}=\frac{\gamma_1}{2\pi\hbar^2}\int_{U}^{\mu}\sqrt{\frac{\varepsilon+
U}{\varepsilon- U}} d\varepsilon\\ \nonumber
=\frac{n_0U}{2\gamma_1}[\sqrt{x^2-1}\pm\ln{(x+\sqrt{x^2-1})}]\,,
\end{eqnarray} where
$n_0=\gamma_1^2/\pi\hbar^2v^2=1.2\times 10^{13}$ cm$^{-2}$ and
$x=\mu/U$.

In order to find the chemical potential $\mu$ and the gap $2|U|$
at the given gate voltage, we minimize  the total energy
containing both the  energy $V$ of the carriers and the energy
$V_f$   of the electrostatic field. Instead of the chemical
potential, it is convenient to use the variable $x$ along with
$U$. Electrons in the $\varepsilon_2$ band or  holes in the
$\varepsilon_3$ band contribute in the total energy of the system
the energy
 \begin{eqnarray} \nonumber\label{wk}
V=\frac{2}{\pi\hbar^2 v^2}\int \varepsilon_{2}(q)qdq=\\
\frac{n_0 U^2}{2\gamma_1} [x\sqrt{x^2-1}+\ln{(x+\sqrt{x^2-1})}]\,.
\end{eqnarray}
 The energy of the electrostatic field,
\begin{equation}\label{wf}
V_f=\frac{1}{8\pi}(dE^2+\epsilon_w d_w E^2_w)
\end{equation}
 can be  written in
terms of the carrier concentrations with the help of relations
(see Fig. \ref{bifig1}):
\begin{equation}\label{ne}
4\pi e(n_1-n_d/2)=E\,  \text{ and}\,4\pi e(n-n_d)=\epsilon_w
E_w\,,
\end{equation}
where $\epsilon_w$ is the dielectric constant of the wafer, the
negative (positive) $n_d$ is the acceptor (donor) concentration
and we suppose that the donors/acceptors are equally divided
between two layers.

 We seek the minimum of the total energy
 as a function of two variables, $U$ and $x$,
 under  the gate  bias  constraint
\begin{equation}
eV_g=-e^2dE-e^2d_wE_w\,.
\end{equation}
Excluding the Lagrange multiplier and assuming the interlayer
distance to be much less than  the
 thickness  $d_w$ of the dielectric wafer, we obtain
the following equation:
\begin{equation}\label{var}
4\pi
e^2d\left(n_2-\frac{n_d}{2}\right)\left(\frac{n_{1x}}{n_x}-\frac{n_{1u}}{n_u}\right)=
\frac{V_x}{n_x}-\frac{V_u}{n_u}\,.
\end{equation}

Let us emphasize, that this equation  is invariant only with
respect to  the simultaneous sign change  in $n_{1,2}$ and $n_d$,
that expresses the charge invariance of the problem. At the fixed
sign of the external doping $n_d$, the gap on the electron and
hole sides of the gate bias is not symmetrical.

The derivatives in Eq. (\ref{var}) are calculated with the help of
Eqs. (\ref{n})--(\ref{ne}). As a result, Eq. (\ref{var}) becomes
\begin{eqnarray}\label{var1}
\frac{\gamma_1 n_d}{Un_0}=\sqrt{x^2-1}\pm\left\{f(x)+\frac{x
f(x)}{\Lambda[x f(x)-\sqrt{x^2-1}]}\right\}
\end{eqnarray}
with the function $f(x)=\ln{(x+\sqrt{x^2-1})}$  and the
dimensionless screening constant
$$\Lambda=\frac{e^2\gamma_1d}{(\hbar v)^2}\,.$$
  For the parameters of graphene $d=3.35\, \AA\,, \gamma_1=0.4$ eV, and
$v=10^8$ cm/s, we get $\Lambda=0.447$.

First, consider  an ideal undoped  bilayer  with $n_d=0$, namely,
$|\gamma_1 n_d/Un_0|\ll 1$. We obtain a solution, as $x_0=6.2784$,
only for one sign in Eq. (\ref{var1}) determining the polarity of
the layers [see Eq. (\ref{nn})]. This value gives
$2|U/\mu|=2/x_0=0.3186$ for the ratio of the gap  to the chemical
potential. According to Eq. (\ref{n}), the gap as a function of
the carrier concentration takes a very simple form:
\begin{equation} \label{w}
 2|U/n|=\frac{2\gamma_1}{n_0\sqrt{x_0^2-1}}=1.08\times10^{-11}
\text{meV} \cdot \text{cm}^2\,,
\end{equation}
where the right-hand side does not depend on the gate bias at all,
but only on the screening constant $\Lambda$.
\begin{figure}[h]
\resizebox{.5\textwidth}{!}{\includegraphics{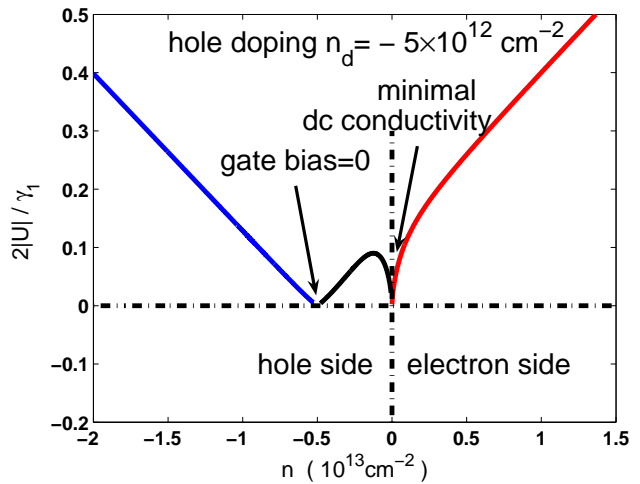}}
\caption{The gap in units of $\gamma_1=0.4 $ eV versus the carrier
concentration for the hole doping with concentration
$n_d=-5\times10^{12}$ cm$^{-2}$; the positive (negative) values of
$n$ correspond to the electron (hole) conductivity.}
\label{bifig3}
\end{figure}
We can compare Eq. (\ref{w}) with the corresponding result of Ref.
\cite{Mc}:
\begin{equation}\label{w1}
2|U/n|=\frac{e^2d}{2\epsilon_0}\left[1+2\Lambda\frac{|n|}{n_0}+
\Lambda\ln\frac{n_0}{|n|}\right]^{-1}\,.
\end{equation}
Both equations are in the numerical agreement at $|n|= 0.1
n_0\simeq 10^{12}$ cm$^{-2}$. However, contrary to Eq. (\ref{w}),
Eq. (\ref{w1}) contains the carrier concentration  in the
right-hand side giving rise to the  more rapid increase in the gap
with $|n|\ll n_0$. This increase also contradicts to the DFT
calculations \cite{GLS}.

 For the bilayer with acceptors or donors,
$n_d\ne0$,  Eq. (\ref{var1}) presents a solution
$w=Un_0/\gamma_1n_d$  as a function of $x$. We obtain, evidently,
the large values of $w$  for $x$ close to $x_0=6.28$.  In this
region of the relatively large $|U|$, we find again  with the help
of Eqs. (\ref{n}) and (\ref{var1})  the linear dependence
\begin{eqnarray}
\nonumber 2|U|=2\gamma_1\frac{|n-n_d|}{n_0x_0}=1.08\,|n-n_d|
\times10^{-11} \text{ meV}\cdot \text{cm}^2.
\nonumber\end{eqnarray}

For the small gap, $|Un_0/\gamma_1n_d|<1$, we obtain different
results for the electron and hole types of conductivity. For
instance, if the bilayer contains acceptors (see Fig.
\ref{bifig3}) with concentration $n_d$, the gap decreases linearly
with the hole concentration and vanishes, when the gate bias is
not applied and the hole concentration equals $n_d$. Starting from
this point, the gap
  increases and, thereafter, becomes  again small
  (equal to zero in Fig. \ref{bifig3}) at the carrier concentration
  corresponding to the minimal value of the dc conductivity.
Therefore, the  difference observed in Ref. \cite{MLS} between
these two values of carrier concentrations, at the zero bias and
at the minimal conductivity, gives directly the donor/acceptor
concentration in the bilayer. Then, for the gate bias applied in
order to increase the electron concentration, the gap is rapidly
opening with the electron appearance.

 We see, that the asymmetry arises
between the electron and hole sides of the gate bias. This
asymmetry can simulate a result of  the hopping integral $\Delta$
in the electron spectrum \cite{CNM}. In order to obtain the gap
dependence for the case of electron doping, $n_d>0$, the
reflection transformation $n\rightarrow -n$ has to be made in Fig.
\ref{bifig3}.

The gap in the vicinity of the minimal conductivity point reaches
indeed a finite value due to several reasons. One of them is the
form of the "mexican hat" shown in Fig. \ref{dis}. Second, the
trigonal warping is substantial at low carrier concentrations.
Finally, the graphene electron spectrum is unstable with respect
to the Coulomb interaction at the low momentum values. For the
graphene monolayer as shown in Ref. \cite{Mi}, the logarithmic
corrections appear at the small momentum. In the case of the
bilayer, the electron self-energy contains the linear corrections,
as can be found using the perturbation theory. The similar linear
terms resulting in a nematic order were also obtained  in the
framework of the renormalization group \cite{VY}.

{\it In conclusions,} the gap $2U$ opening in the gated graphene
bilayer has an intriguing behavior as a function of carrier
concentration. In the presence of the external doping charge,
i.\,e. donors or acceptors, this function is asymmetric on the
hole and electron sides of the gate bias and it is linear only
for the large gate bias.

This work was supported by the Russian Foundation for Basic
Research (grant No. 07-02-00571). The author is grateful to the
Max Planck Institute for the Physics of Complex Systems for
hospitality in Dresden.

\end{document}